\documentclass[reprint,
 amsmath,amssymb,
 aps,
pre,
floatfix,
]{revtex4-2}

\usepackage{graphicx}
\usepackage{dcolumn}
\usepackage{bm}
\usepackage{xcolor}
\usepackage{ulem}

\begin{document}

\title{Synchronization and metachronal waves in an array of eukaryotic flagella}

\author{Yukinori Wakahara}
\affiliation{Department of Physics, Tohoku University, Sendai 980-8578, Japan}
\author{Nariya Uchida}%
 \email{nariya.uchida@tohoku.ac.jp}
 \affiliation{Department of Physics, Tohoku University, Sendai 980-8578, Japan}

\date{\today}

\begin{abstract}
We investigate synchronization and metachronal-wave formation in a one-dimensional array of eukaryotic flagella using an elastohydrodynamic model. In contrast to a two-flagellum system, where only in-phase synchronization is stable, larger arrays are found to support stable metachronal waves with finite phase differences. Direct numerical simulations show that metachronal waves appear with increasing probability as the number of flagella increases. To explain this many-body effect, we construct a phase description for the array from that of the pair problem and analyze the stability of phase-locked states with nearest-neighbor hydrodynamic coupling. The analysis shows that increasing system size enlarges the set of stable phase-locked modes, thereby promoting metachronal-wave selection. A continuum description further relates these collective states to advection and diffusion of the phase-difference field. These results provide a simple theoretical framework for understanding how hydrodynamic interactions generate robust metachronal waves in flagellar arrays.
\end{abstract}

\maketitle


\section{INTRODUCTION}

Cilia and flagella are ubiquitous organelles that generate fluid flow and propulsion in a wide range of biological systems. They play essential roles in processes such as the swimming of microorganisms~\cite{tamm1972ciliary,brumley2012hydrodynamic} and fluid transport in multicellular organisms, including mucus clearance in the respiratory tract~\cite{sleigh1988propulsion,lever2024metachrony} and cerebrospinal fluid circulation~\cite{faubel2016cilia}. In many cases, large numbers of cilia or flagella are densely arranged on a surface, and their collective motion gives rise to coordinated spatiotemporal patterns~\cite{wan2024mechanisms}.

A prominent example of such collective behavior is the emergence of metachronal waves, in which neighboring cilia or flagella beat with a constant phase difference, forming traveling waves along the array. Metachronal waves have been widely observed in both natural systems and experiments~\cite{tamm1972ciliary,brumley2012hydrodynamic,ringers2023novel}, and are believed to enhance transport efficiency by generating directional fluid flow~\cite{elgeti2013emergence,osterman2011finding,lever2024metachrony}. Understanding the physical mechanisms underlying the formation and stability of these waves is therefore a fundamental problem in active matter and biological physics.

To elucidate the origin of synchronization and metachronal waves, a variety of theoretical models have been proposed. One class of approaches is based on minimal phase oscillator descriptions, often referred to as rotor models, in which each cilium is represented as a driven oscillator moving along a prescribed trajectory~\cite{vilfan2006hydrodynamic,elfring2009hydrodynamic,uchida2011generic}. These models have demonstrated that hydrodynamic interactions alone can lead to phase locking and the emergence of metachronal waves under appropriate conditions, such as broken symmetry in the driving force or geometric anisotropy~\cite{niedermayer2008synchronization,meng2021conditions}.

Another class of models explicitly incorporates the internal driving mechanism of cilia and flagella, namely the activity of molecular motors such as dynein. In these molecular motor-based models, the beating pattern arises from the interplay between active internal forces and the elasticity of the filament~\cite{brokaw1972computer,camalet2000generic,lindemann1994geometric,riedel2007molecular,bayly2014equations}. Such models can capture waveform generation and provide a more detailed description of the oscillatory dynamics.

While these two approaches have provided important insights, they represent different levels of description. Rotor models offer a coarse-grained and analytically tractable framework for studying phase synchronization, but neglect the underlying waveform dynamics. In contrast, molecular motor-based models capture the internal mechanics of individual flagella, but are often complex and less amenable to analytical treatment in large systems. In particular, although recent studies have clarified several mechanisms for metachronal coordination in extended ciliary arrays~\cite{elgeti2013emergence,meng2021conditions,chakrabarti2022multiscale,ringers2023novel}, the mechanisms by which metachronal waves are selected and stabilized in spatially extended arrays remain incompletely understood within a unified framework.

In this study, we investigate synchronization and metachronal wave formation using an intermediate, mesoscale description based on an elastohydrodynamic model of flagellar beating. This model, originally developed to describe a pair of eukaryotic flagella~\cite{goldstein2016elastohydrodynamic}, incorporates both the elastic deformation of the filament and its hydrodynamic interaction with the surrounding fluid, while remaining sufficiently simple for analytical treatment. In this framework, the near-field hydrodynamic interaction between nearby slender filaments can be approximated analytically in the regime of small interfilament spacing compared with the filament length~\cite{man2016hydrodynamic}. Moreover, Kawamura and Tsubaki~\cite{kawamura2018phase} formulated a phase-reduction theory for this pair model and showed that a pair of adjacent flagella exhibits stable in-phase synchronization through hydrodynamic interactions.

By extending this framework to a one-dimensional array of flagella, we aim to bridge the gap between minimal phase oscillator models and detailed molecular descriptions.
Recent work by Jung \textit{et al.}~\cite{jung2025emergence} demonstrated that metachronal waves can arise in a chain of symmetrically beating filaments when a tilt-induced geometric misalignment breaks the left-right symmetry and generates a constant time lag between adjacent filaments. In contrast, we show that metachronal-wave states can emerge through spontaneous symmetry breaking even without imposed geometric asymmetry, and that their selection becomes increasingly likely in larger arrays.

To elucidate this mechanism, we combine direct numerical simulations with a phase reduction theory and show how spatial extension and hydrodynamic interactions influence the emergence and stability of synchronized states. In particular, we show that, although a pair of flagella exhibits only in-phase synchronization, a spatially extended array supports a broader class of stable phase-locked states, including metachronal waves. Furthermore, we demonstrate that the selection of these states depends strongly on the system size and can be understood in terms of the stability structure of the phase dynamics.

\begin{figure}
\includegraphics[width=0.6\columnwidth]{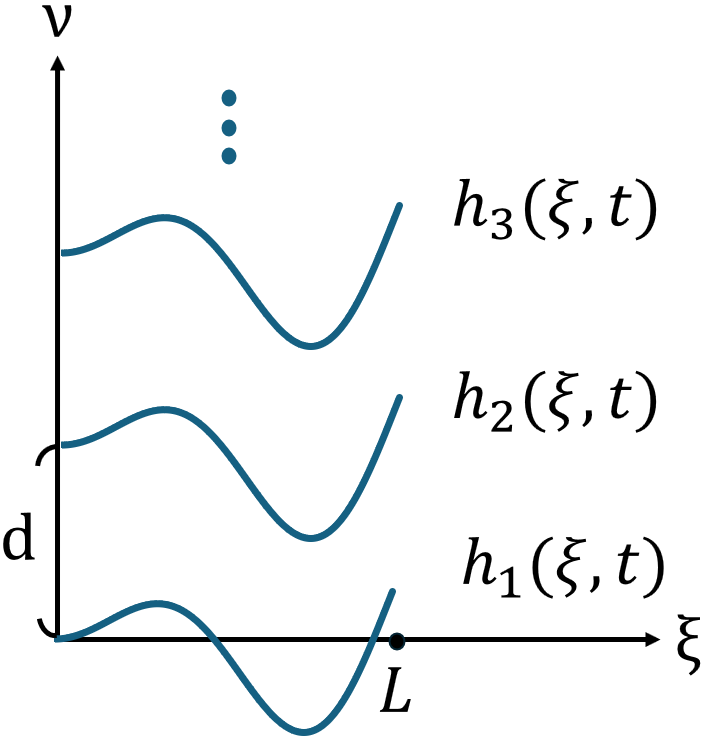}
\caption{
Schematic of the coordinate system for a one-dimensional array of flagella. The flagella are aligned along the $\nu$ direction, each flagellum has length $L$, and the spacing between adjacent flagella is $d$. The transverse displacement of the $j$th flagellum is denoted by $h_j(\xi,t)$.
\label{fig:set_up}}
\end{figure}

\section{MODEL}

In this study, we extend the elastohydrodynamic model of Ref.~\cite{goldstein2016elastohydrodynamic}, which describes the motion of two eukaryotic flagella, to a one-dimensional flagellar array. As shown in Fig.~\ref{fig:set_up}, we consider a one-dimensional array of $N_c$ flagella embedded in three-dimensional space. The flagella are aligned in the $\nu$ direction and are described on a two-dimensional coordinate system $(\xi,\nu)$ in nondimensional units. The flagella are labeled sequentially from bottom to top as flagellum 1, flagellum 2, $\ldots$, and flagellum $N_c$. Let $h_j(\xi,t)$ denote the transverse displacement of the $j$th flagellum at position $\xi$ and time $t$. We consider the small-deformation regime, where the transverse displacement $h_j(\xi,t)$ is sufficiently small and the flagella can be approximated as nearly straight. Each flagellum has length $L$, and the spacing between adjacent flagella is $d$. Under the assumption of nearest-neighbor hydrodynamic coupling, the equation of motion for flagellum $j$ $(j=1,\ldots,N_c)$ can be written as follows: 
\begin{flalign}
\frac{\partial h_j}{\partial t}
&= - c\frac{\partial h_j}{\partial \xi}
 - 2\frac{\partial^2 h_j}{\partial \xi^2}
 - \frac{\partial^4 h_j}{\partial \xi^4}
 + \left(\frac{\partial^2 h_j}{\partial \xi^2} \right)^3  \notag\\
&\quad
 + \varepsilon \left(
 \frac{\partial h_{j-1}}{\partial t}
 + \frac{\partial h_{j+1}}{\partial t}
 \right).
\label{eq:multi-flagella-a}
\end{flalign}
The left-hand side represents the viscous drag exerted by the surrounding fluid on flagellum $j$. The first term on the right-hand side is an advective term that propagates bending waves to the right, where $c$ sets the propagation speed. The second term represents an effective negative tension that induces self-sustained oscillations of flagellum $j$, and the third term accounts for bending elasticity. The fourth term is a nonlinear term that saturates the growth of the oscillation amplitude. The last term represents hydrodynamic interactions with the neighboring flagella $j\pm1$. Note that, for open boundary conditions,
the interaction term remains only for one neighbor of the flagellum. We consider nearest-neighbor coupling as a minimal model to isolate the essential mechanism of metachronal-wave formation in the array.

The parameter $\varepsilon$ characterizes the strength of the hydrodynamic coupling. If $a$ denotes the radius of a flagellum, then $\varepsilon$ can be written as $\varepsilon = \ln(L/d)/\ln(L/a)$ in the asymptotic limit $d \ll L$. The validity of this approximation has been examined in previous work, where good agreement with slender-body theory was reported for parameters such as $d/L = 0.1$ and $a/L = 0.025$~\cite{man2016hydrodynamic}. 

For each flagellum, we impose hinged boundary conditions at the left end,
\begin{equation}
h_j(0,t) = 0, \quad \partial_{\xi\xi} h_j(0,t) = 0,
\end{equation}
and free boundary conditions at the right end,
\begin{equation}
\partial_{\xi\xi} h_j(L,t) = 0, \quad \partial_{\xi\xi\xi} h_j(L,t) = 0.
\end{equation}
Here, $\partial_\xi$ denotes differentiation with respect to $\xi$.
All variables are nondimensionalized following the procedure of Ref.~\cite{goldstein2016elastohydrodynamic}.

\begin{figure*}
\centering

\begin{minipage}{\columnwidth}
    \centering
    \includegraphics[width=\linewidth]{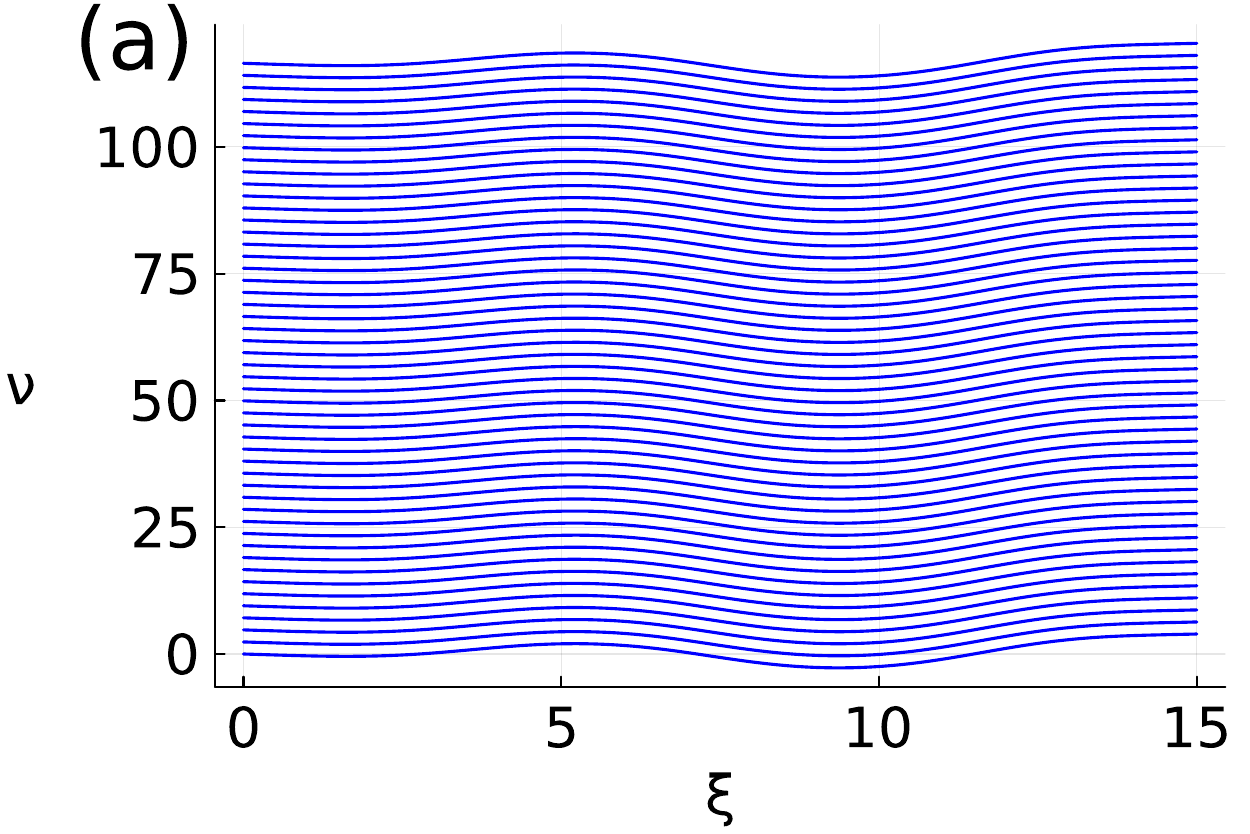}
\end{minipage}
\hfill
\begin{minipage}{\columnwidth}
    \centering
    \includegraphics[width=\linewidth]{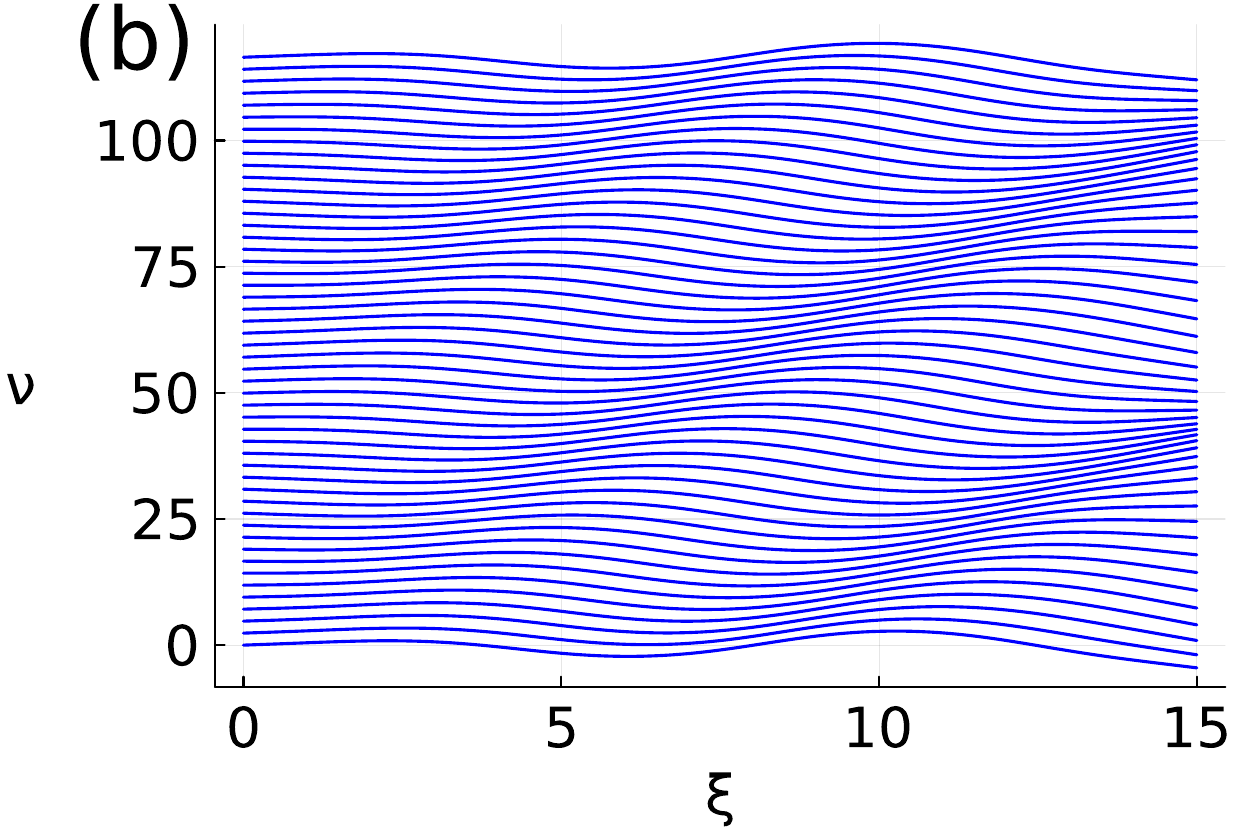}
\end{minipage}
\hfill
\begin{minipage}{\columnwidth}
    \centering
    \includegraphics[width=\linewidth]{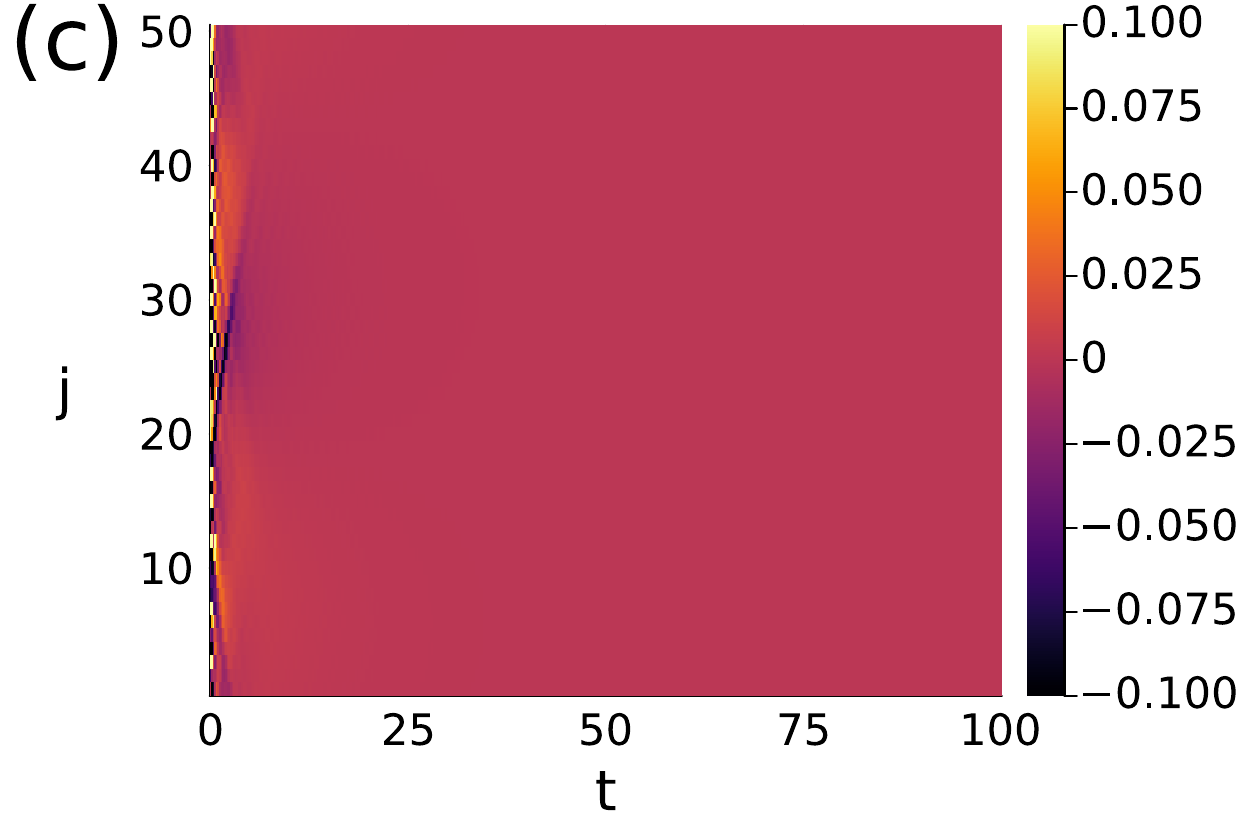}
\end{minipage}
\hfill
\begin{minipage}{\columnwidth}
    \centering
    \includegraphics[width=\linewidth]{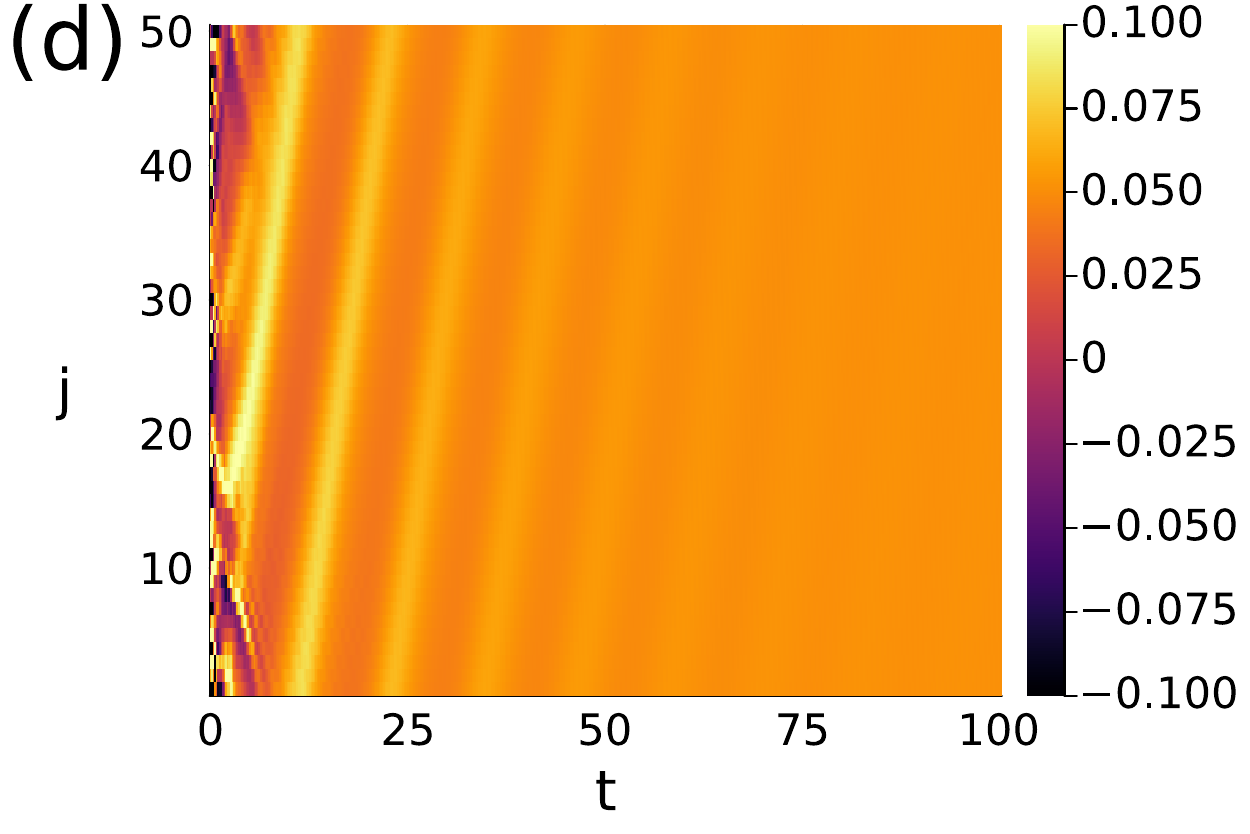}
\end{minipage}
\caption{
(a, b) Snapshots of the flagellar array at nondimensional time $t=40$ under periodic boundary conditions for $N_c=50$, $c=1.5$, and $\varepsilon=0.40$. (a) In-phase synchronized state. (b) Metachronal-wave state. (c,d) Time evolution of the time delay $\tau_j(t)$ between neighboring flagella for the states shown in (a) and (b), respectively. In (c) and (d), the horizontal axis represents time $t$, the vertical axis denotes the flagellum index $j$, and the color indicates the value of $\tau_j(t)$.
}
\label{fig:synchronized}
\end{figure*}

\section{NUMERICAL RESULTS}

In this section, we present numerical results for the one-dimensional flagellar array under both periodic and open boundary conditions.

Equation~(1) for the one-dimensional flagellar array was solved numerically using an explicit Euler method with a fixed time step. Spatial derivatives were evaluated using finite-difference schemes. The advective term was discretized using a first-order one-sided finite-difference scheme. The spatial and temporal step sizes were set to \(\Delta x=0.15\) and \(\Delta t=10^{-5}\), respectively. The advection parameter was chosen as \(c=1.5\), and the hydrodynamic coupling strength was set to \(\varepsilon=0.4\).

As an initial condition satisfying the boundary conditions, we impose
\begin{equation}
h_j(\xi,0)
=
A\,\xi^3(\xi-L)^4
\sin\left(
\frac{4\pi \xi}{L}+\theta_j^0
\right).
\end{equation}
Here, $A$ denotes the amplitude and is set to $10^{-8}$. The phase $\theta_j^0$ is drawn independently from a uniform distribution on $[0,2\pi)$. The initial amplitude of the flagella is chosen to be much smaller than that in the steady state. The amplitude subsequently grows in time and eventually reaches a stationary value.

Denoting the number of flagella by $N_c$, we solved the equations of motion for $N_c=10$, $30$, $50$, $70$ and $100$. In the $\nu$ direction, we imposed either periodic or open boundary conditions.

To quantify synchronization, we define the time delay $\tau_j(t)$ between neighboring flagella $j$ and $j+1$ at time $t$ as the time shift that minimizes
\begin{equation}
\int_0^L d\xi \,
\left[
h_{j+1}(\xi,t)-h_j(\xi,t+\tau_j(t))
\right]^2 .
\end{equation}
We further define the system-averaged time delay as
\begin{equation}
\tau(t)=\langle \tau_j(t)\rangle_j ,
\end{equation}
and denote by $\tau_s$ its steady-state value.

\begin{figure*}
\centering
\begin{minipage}{\columnwidth}
    \centering
    \includegraphics[width=0.9\linewidth]{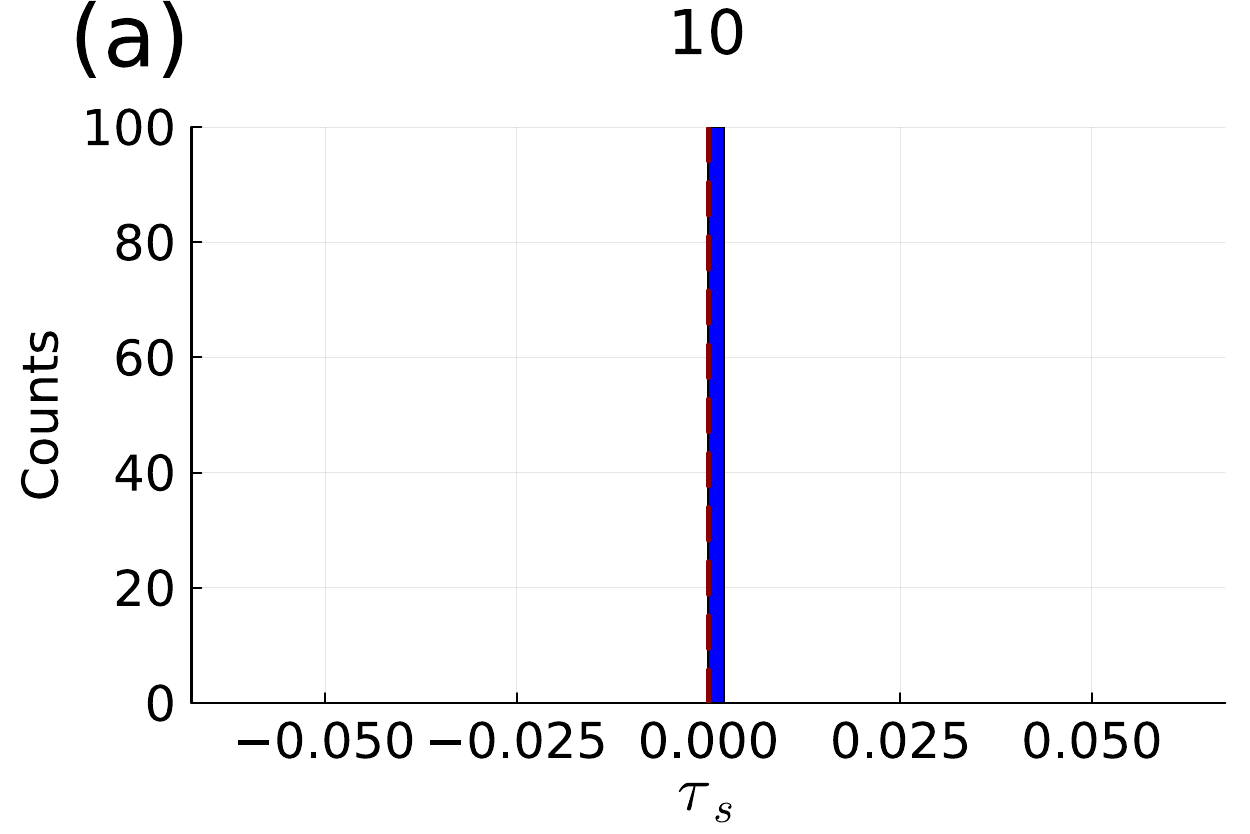}
\end{minipage}
\hfill
\begin{minipage}{\columnwidth}
    \centering
    \includegraphics[width=0.9\linewidth]{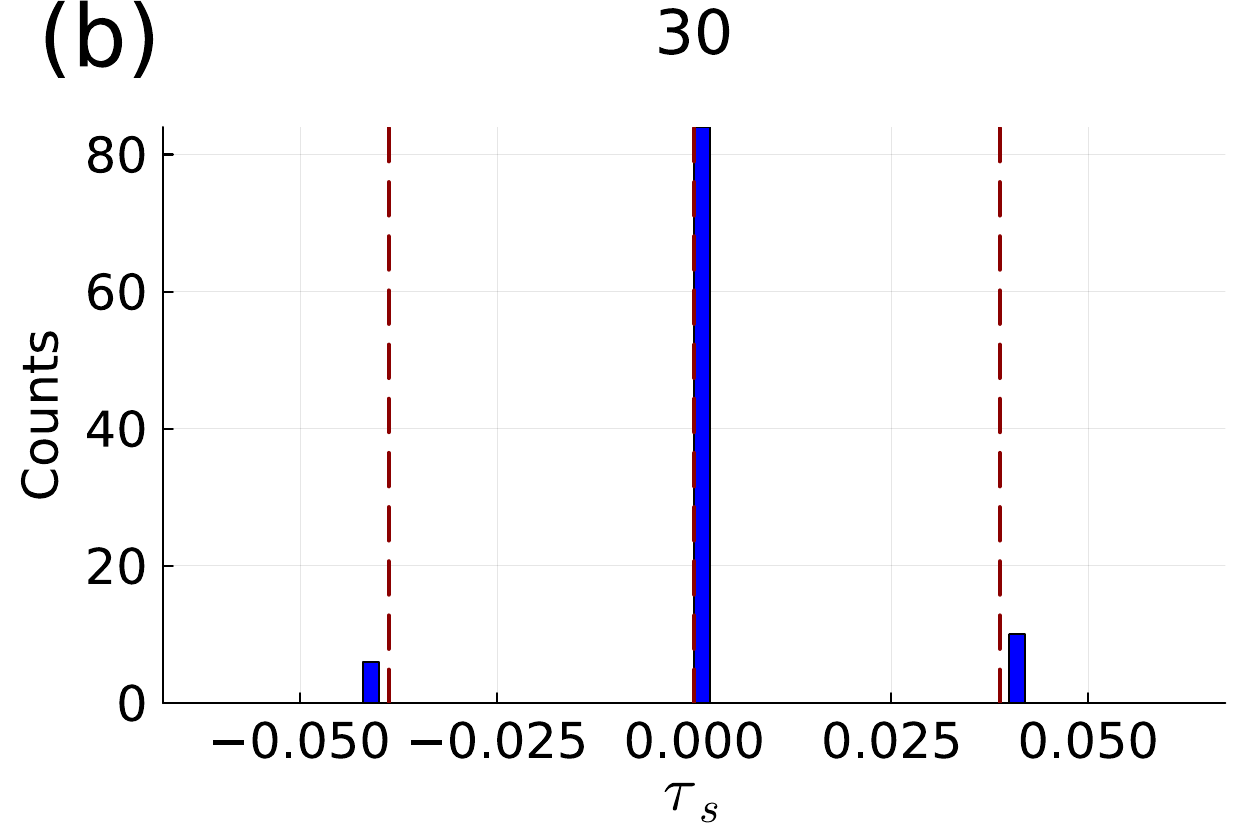}
\end{minipage}
\hfill
\begin{minipage}{\columnwidth}
    \centering
    \includegraphics[width=0.9\linewidth]{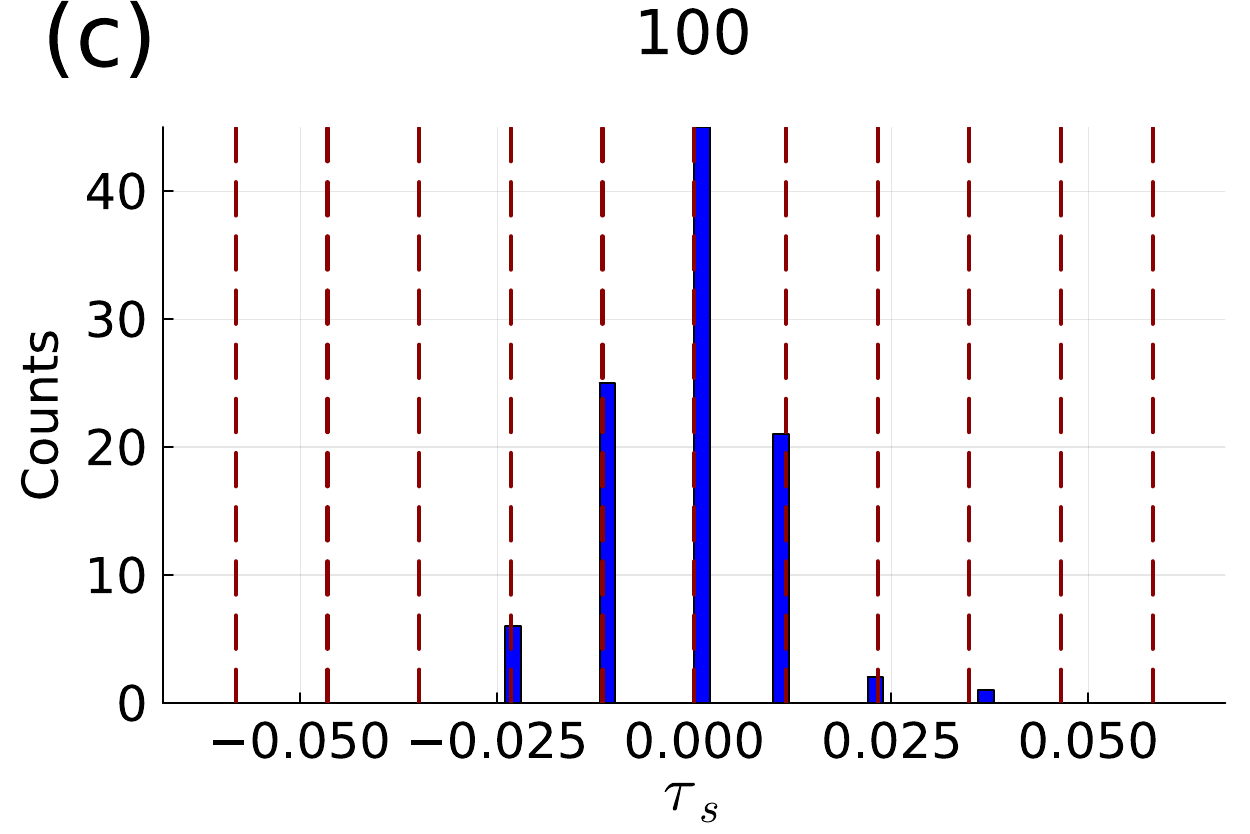}
\end{minipage}
\hfill
\begin{minipage}{\columnwidth}
    \centering
    \includegraphics[width=0.9\linewidth]{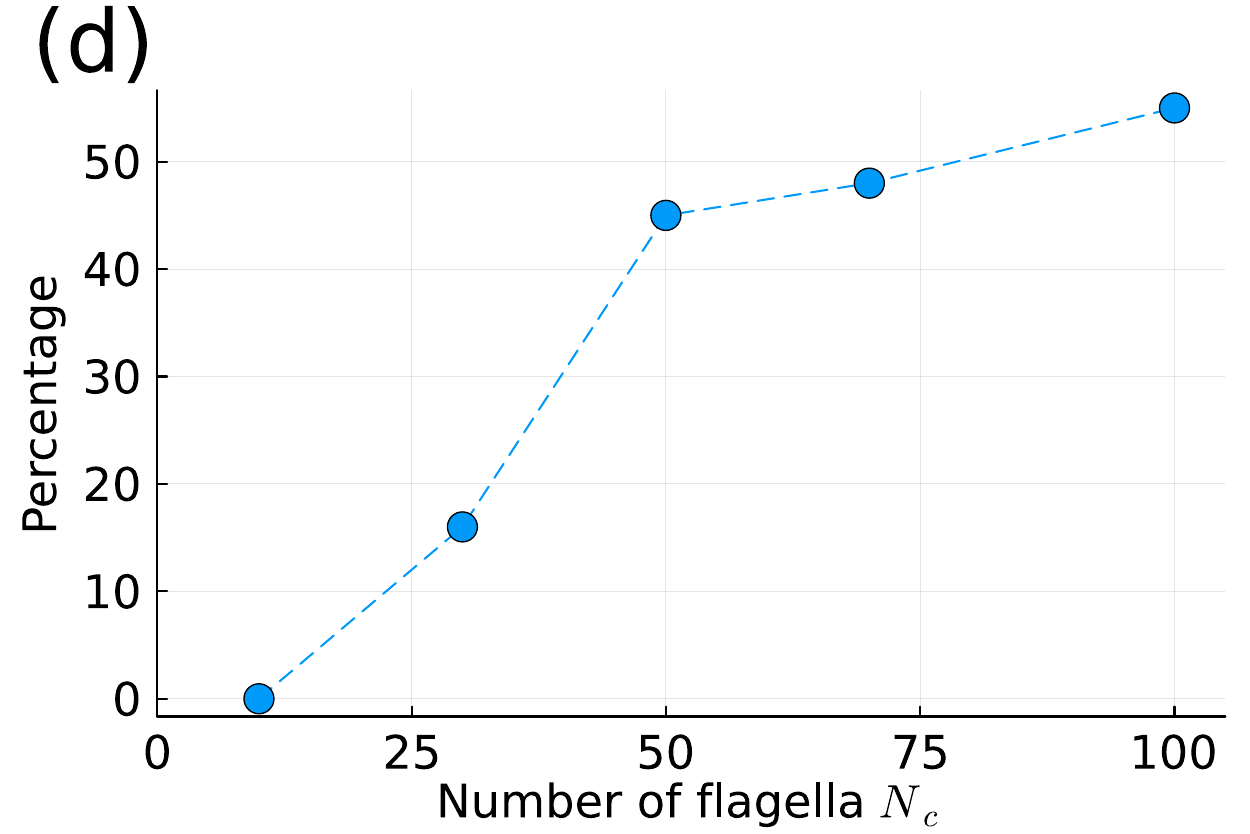}
\end{minipage}
\caption{
(a-c) Histograms of the steady-state system-averaged time delay $\tau_{\mathrm{s}}$ under periodic boundary conditions with 100 samples. Dashed lines indicate the analytical values of $\tau_s$ predicted by Eq.~(\ref{eq:p-tau}) with $\omega = 5.4$. 
(a) $N_c=10$. Only the in-phase synchronized state with $\tau_{\mathrm{s}}=0$ is observed. 
(b) $N_c=30$. In addition to the in-phase synchronized state, metachronal-wave states with $\tau_{\mathrm{s}} \neq 0$ appear. 
(c) $N_c=100$. Compared with the case $N_c=30$, the peak at $\tau_{\mathrm{s}}=0$ is lower, indicating that metachronal-wave states are more likely to be selected. 
(d) Percentage of samples that reached metachronal-wave states among 100
realizations for each \(N_c\). 
}
\label{fig:p-hist}
\end{figure*}

\subsection{Periodic boundary conditions}
\label{subsec:periodic}

We first present numerical results for periodic boundary conditions in the $\nu$ direction. At the initial stage, the flagella beat independently without synchronization. As time evolves, however, hydrodynamic interactions induce phase locking among neighboring flagella, and the system gradually approaches a synchronized state. Depending on the initial condition, the steady state is found to be either an in-phase synchronized state or a metachronal-wave state.

Figure~\ref{fig:synchronized}(a,b) shows representative synchronized states for $N_c=50$ flagella. In Fig.~\ref{fig:synchronized}(a), all flagella oscillate with nearly identical waveforms and phases, corresponding to the in-phase synchronized state. In Fig.~\ref{fig:synchronized}(b), neighboring flagella maintain a constant nonzero phase difference, corresponding to a metachronal wave.
The corresponding time-evolution movies are provided as Supplementary Movies 1 
and 2, respectively, in the Supplemental Material~\cite{SM}.

Figures~\ref{fig:synchronized}(c) and \ref{fig:synchronized}(d) show the time evolution of the local time delay $\tau_j(t)$ for the states shown in Fig.~\ref{fig:synchronized}(a) and Fig.~\ref{fig:synchronized}(b), respectively. In both cases, the time delays are initially random, but they gradually converge to constant values as time evolves. In Fig.~\ref{fig:synchronized}(c), the steady state ($t \gtrsim 5$) shows $\tau_j(t) \approx 0$, indicating in-phase synchronization. In Fig.~\ref{fig:synchronized}(d), the steady state ($t \gtrsim 80$) shows that $\tau_j(t)$ converges to a constant nonzero value, corresponding to a metachronal wave.

In addition, stripe-like structures are observed in Fig.~\ref{fig:synchronized}(d) during the intermediate stage ($5 \lesssim t \lesssim 80$). These stripes indicate that regions with relatively large phase differences propagate along the array. As synchronization develops, the stripe pattern gradually disappears, and the system converges to a steady state with an approximately uniform phase difference.


Next, we investigate how the number of flagella affects the selected synchronized state. For $N_c=10$, $30$, $50$, $70$, and $100$, we computed the steady-state system-averaged time delay $\tau_s$ from 100 independent samples for each case.

Figure~\ref{fig:p-hist} shows histograms of $\tau_s$ obtained from these samples. For $N_c=10$, only the in-phase synchronized state with $\tau_s=0$ is observed. When the number of flagella is increased to $N_c=30$, metachronal-wave states with $\tau_s \neq 0$ appear in addition to the in-phase synchronized state. For $N_c=100$, 
the peak at $\tau_s=0$ is lower than that for $N_c=30$, indicating that metachronal-wave states are more frequently selected. Thus, as the number of flagella increases, metachronal waves are more likely to be realized as the steady state. 
Figure~\ref{fig:p-hist}(d) shows the percentage of metachronal-wave samples among 100 realizations for $N_c = 10,\:30,\:50,\:70,$ and $100$. The percentage increases monotonically with increasing $N_c$.

Since the present model does not include any directional bias along the array, metachronal waves emerge through spontaneous symmetry breaking. Statistically, the positive and negative values of $\tau_s$ appear with approximately equal probability. The discrete distribution of $\tau_s$ originates from the periodic boundary conditions, which require that the time delay satisfy 
\begin{equation}
    \label{eq:p-tau}
    \tau_\mathrm{s} = \frac{2\pi n}{\omega N_c},
\end{equation}
where $n$ is an integer and $\omega$ is the angular frequency of the steady flagellar beating.

The angular frequency $\omega$ in the steady state was computed via Fourier transform. As a result, for $N_c = 10, 50, 100$, we obtained $\omega \approx 5.4$ in all cases. This value is approximately five times larger than the intrinsic beating frequency of an isolated flagellum, $\omega = 1.1$. This can be understood from the fact that, in the synchronized state, $h_i(\xi,t) = h_{\mathrm{s}}$ satisfies the following relation, which implies that the characteristic time scale is reduced to one-fifth when $\varepsilon=0.4$.
\begin{equation}
(1-2\varepsilon)\frac{\partial h_{\mathrm{s}}}{\partial t}
= - c\frac{\partial h_{\mathrm{s}}}{\partial \xi}
 - 2\frac{\partial^2 h_{\mathrm{s}}}{\partial \xi^2}
 - \frac{\partial^4 h_{\mathrm{s}}}{\partial \xi^4}
 + \left(\frac{\partial^2 h_{\mathrm{s}}}{\partial \xi^2} \right)^3.
\label{eq:sync-flagella}
\end{equation}
For metachronal waves, if the phase difference is sufficiently small, the above relation also holds to first order in the time delay $\tau$. Therefore, metachronal waves with small phase differences likewise exhibit $\omega \approx 5.4$. However, for metachronal waves with a larger phase difference, corresponding to a time delay of $\tau \simeq 0.05$, the frequency decreases. For example, in a case $N_c = 50$, we obtain $\omega = 4.8$. This reduction can be attributed to the increase in phase difference, which leads to less constructive hydrodynamic interactions among the flagella, thereby reducing the efficiency of collective driving and slowing down the overall oscillatory dynamics.

Setting $\omega = 5.4$ in Eq.~(\ref{eq:p-tau}) yields $\tau_s$, shown by the dashed line in Fig.~\ref{fig:p-hist}, in good agreement with the numerical results.

\begin{figure*}
\centering

\begin{minipage}{\columnwidth}
    \centering
    \includegraphics[width=0.98\linewidth]{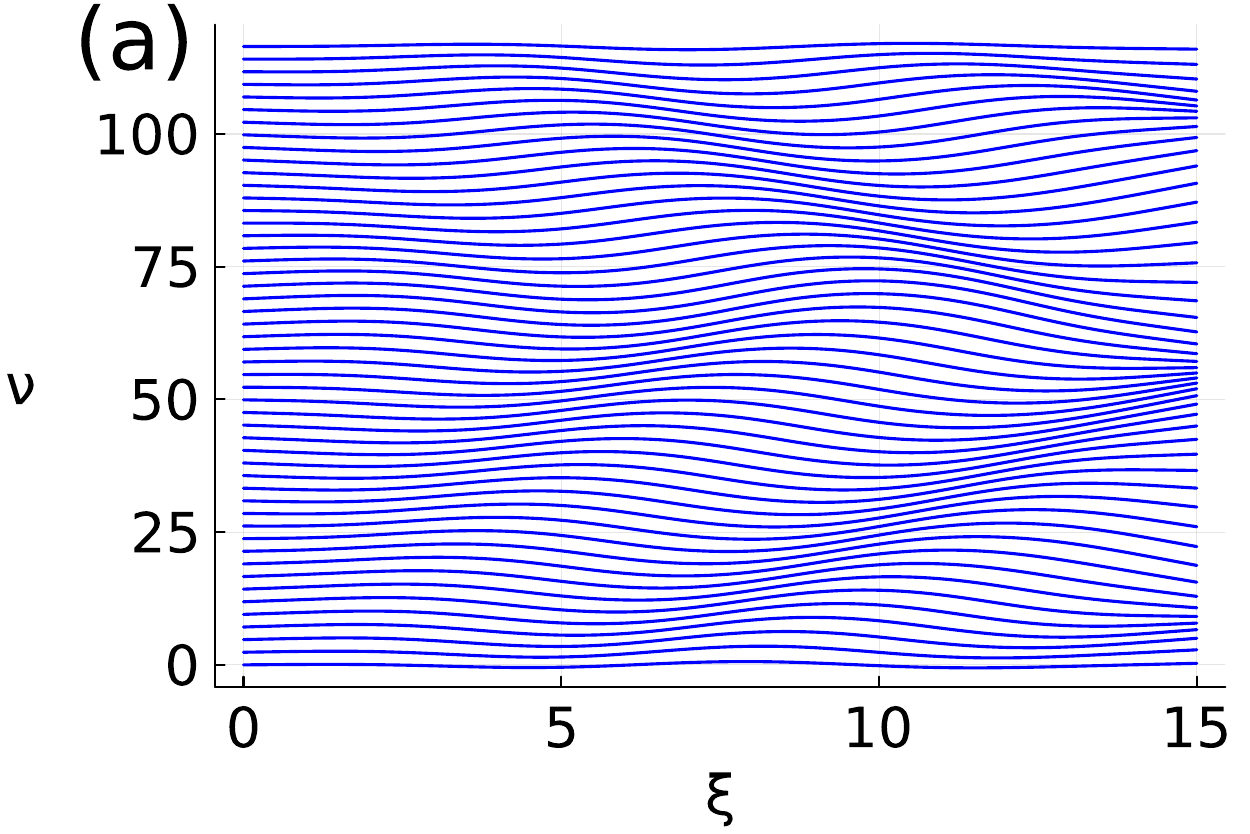}
\end{minipage}
\hfill
\begin{minipage}{\columnwidth}
    \centering
    \includegraphics[width=0.98\linewidth]{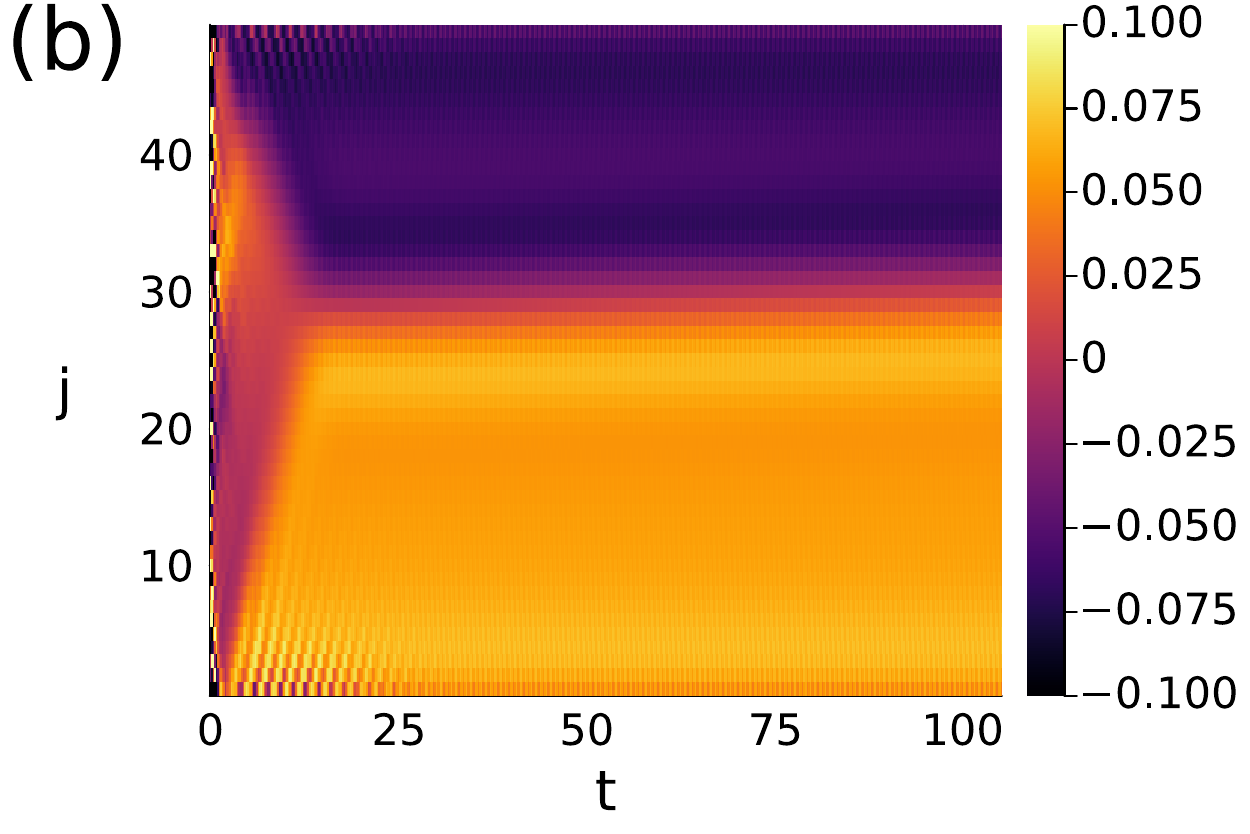}
\end{minipage}
\caption{
(a) Snapshot of the flagellar array at nondimensional time $t=40$ under open boundary conditions for $N_c=50$. A chevron pattern, in which positive and negative $\tau_j(t)$ coexist, emerges around the center of the array. (b) Time evolution of the time delay $\tau_j(t)$ between neighboring flagella for the state shown in (a).
In (b), the horizontal axis represents time \(t\), the vertical axis denotes the
flagellum index \(j\), and the color indicates the value of \(\tau_j(t)\).
}
\label{fig:open}
\end{figure*}


\subsection{Open boundary conditions}
\label{subsec:open}

Next, we consider the case where open boundary conditions are imposed along the $\nu$ direction. In contrast to the periodic case, the steady state is characterized by a chevron pattern rather than a spatially uniform phase-lag state.

Figure~\ref{fig:open}(a) shows a snapshot at nondimensional time $t=40$ of the flagellar waveforms for $N_c=50$. For $\nu \lesssim 70$, the neighboring flagella exhibit positive time delays $\tau_j(t)$, whereas for $\nu \gtrsim 70$, negative $\tau_j(t)$ are observed. As a result, a chevron pattern, in which positive and negative $\tau_j(t)$ coexist, emerges. The sign of $\tau_j(t)$ changes near the center of the array. Such a pattern has also been reported in rotor models with radial flexibility~\cite{niedermayer2008synchronization}.
The corresponding time-evolution movie is provided as Supplementary Movie 3
in the Supplemental Material~\cite{SM}.

Figure~\ref{fig:open}(b) shows a colormap of the time evolution of the time delay $\tau_j(t)$ for the chevron pattern shown in Fig.~\ref{fig:open}(a). The sign of $\tau_j(t)$ changes around $j \approx 25$, forming the chevron pattern. For $t \lesssim 30$, stripe-like structures are observed, indicating the presence of regions where the time delay $\tau_j(t)$ differs from that of the surrounding flagella. As time evolves, the stripe pattern gradually fades, and for $t \gtrsim 30$ the color pattern becomes stationary. Thus, unlike the periodic case, the system does not converge to a spatially uniform phase lag, but instead reaches a steady chevron pattern.

\begin{figure*}
\centering
\begin{minipage}{\columnwidth}
    \centering
    \includegraphics[width=0.9\linewidth]{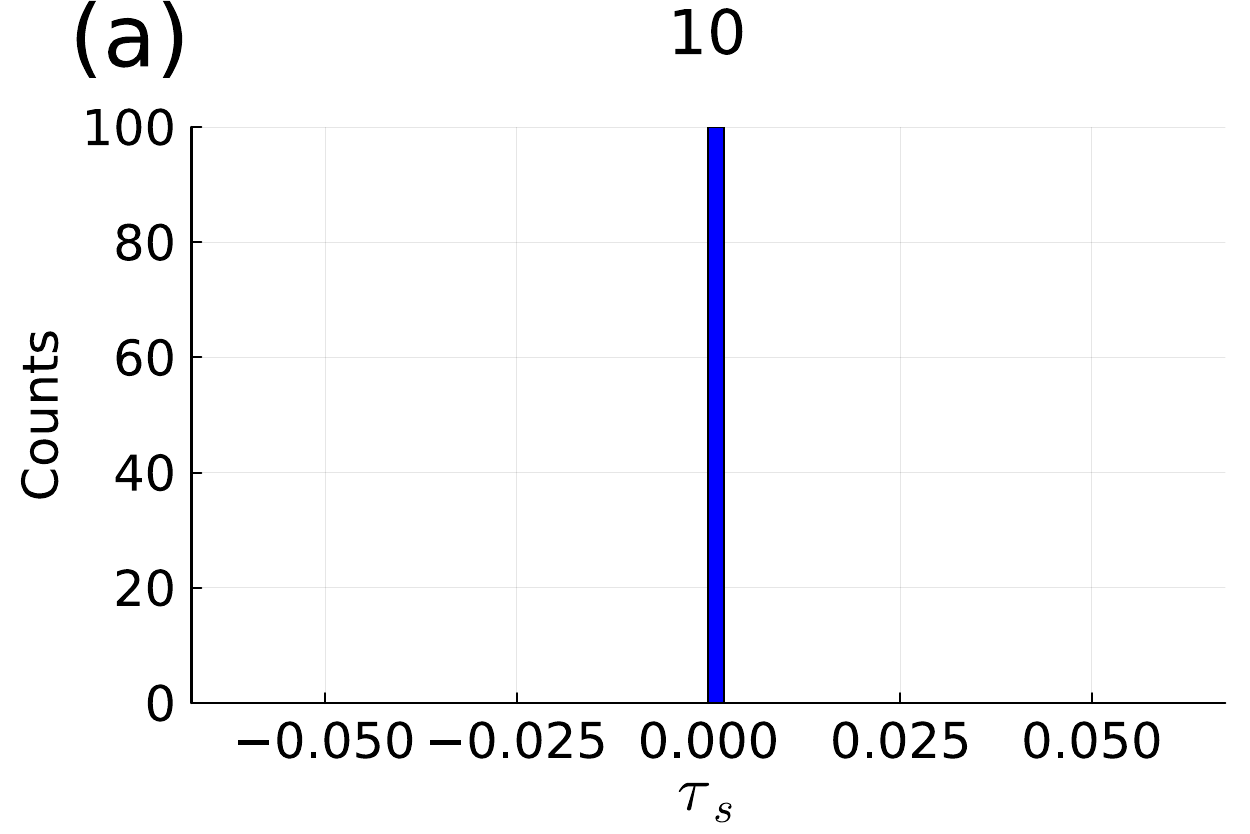}
\end{minipage}
\hfill
\begin{minipage}{\columnwidth}
    \centering
    \includegraphics[width=0.9\linewidth]{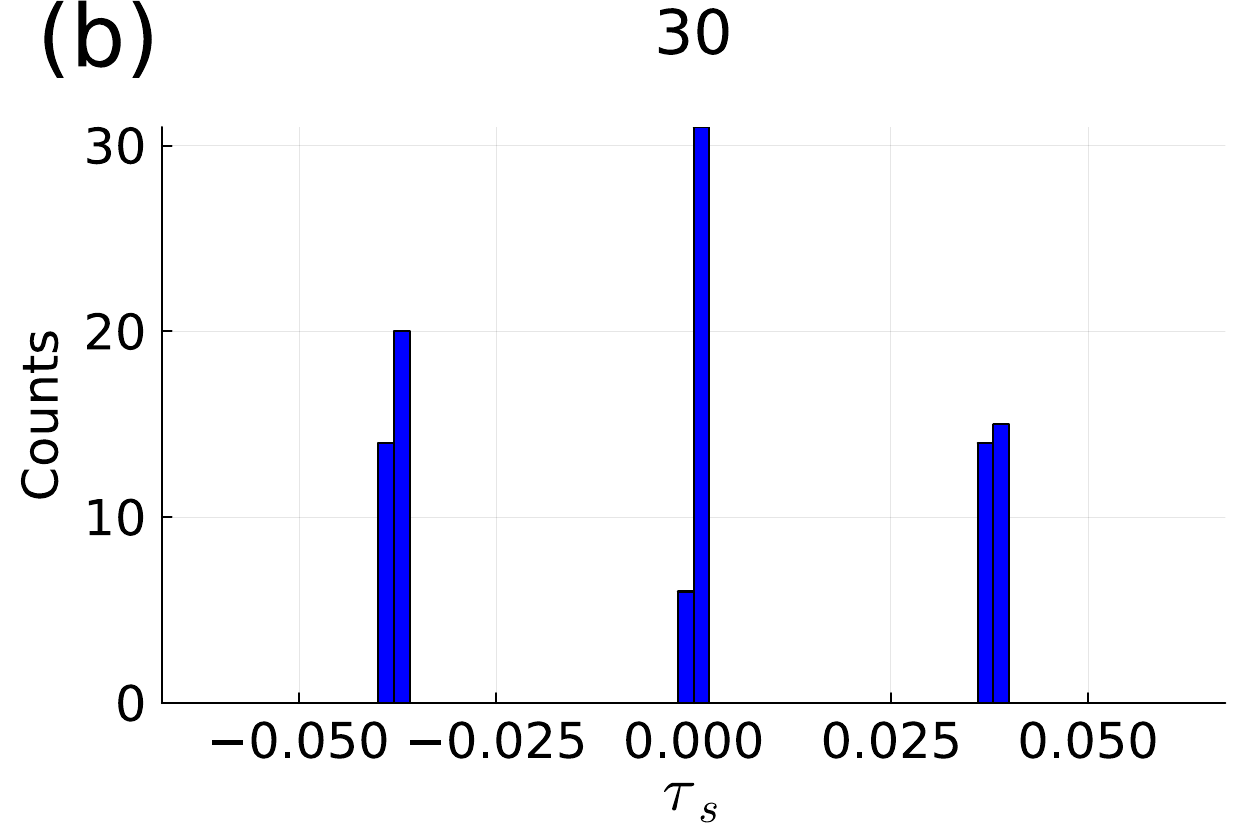}
\end{minipage}
\hfill
\begin{minipage}{\columnwidth}
    \centering
    \includegraphics[width=0.9\linewidth]{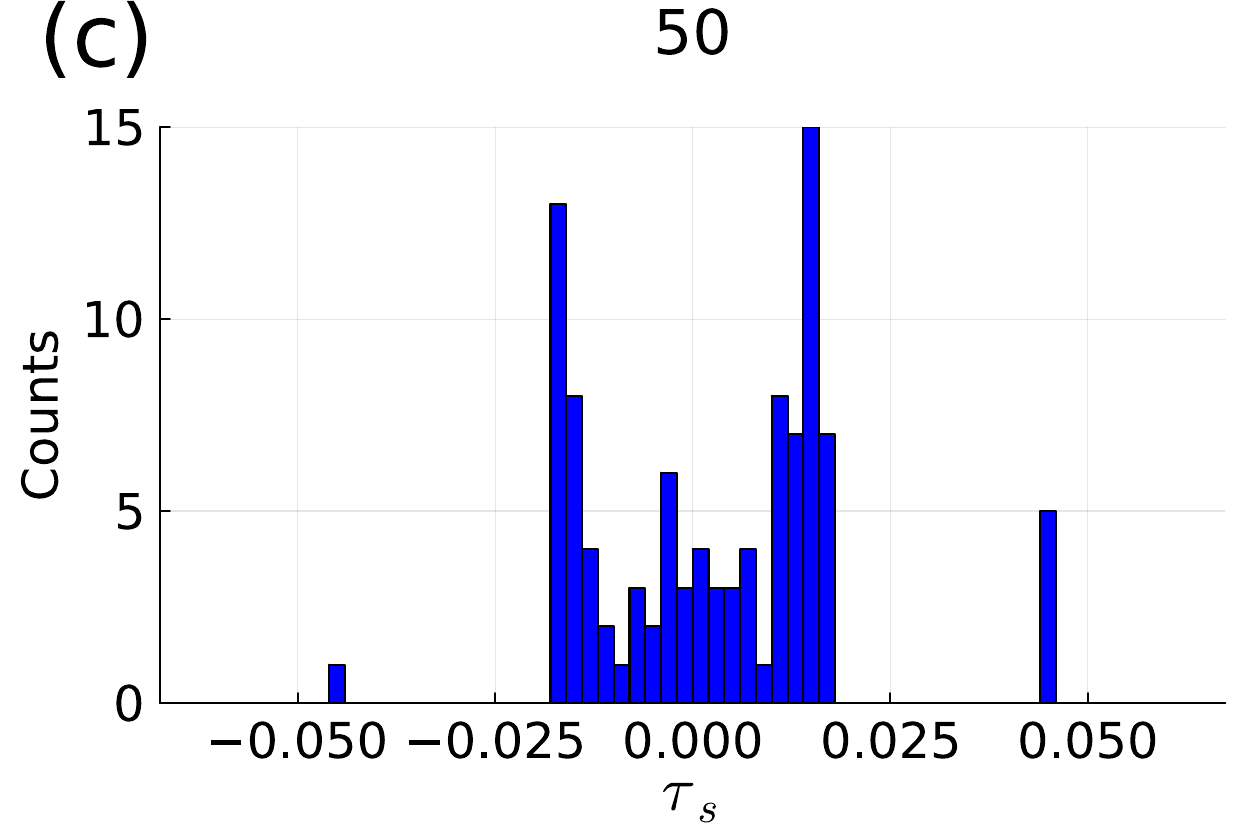}
\end{minipage}
\hfill
\begin{minipage}{\columnwidth}
    \centering
    \includegraphics[width=0.9\linewidth]{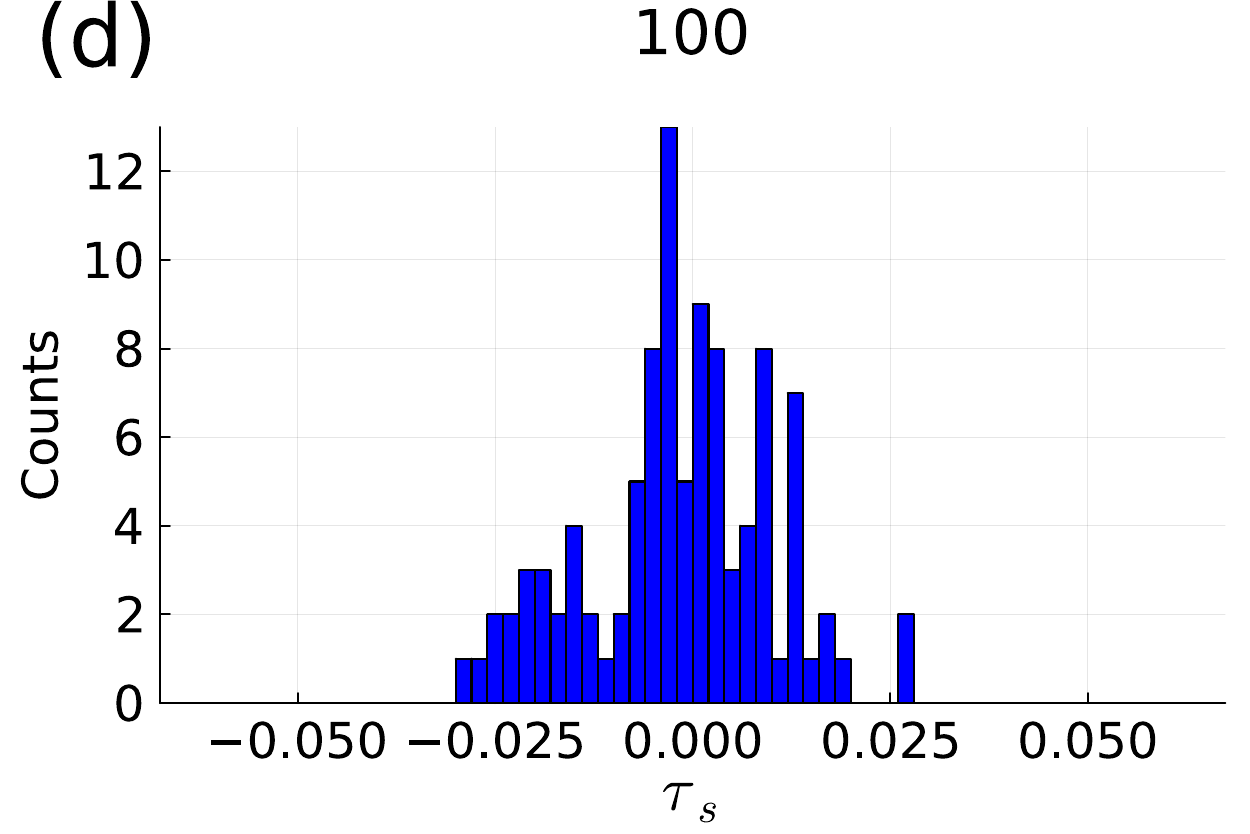}
\end{minipage}
\caption{
Histograms of the steady-state system-averaged time delay $\tau_{\mathrm{s}}$ under open boundary conditions. In all samples, the steady state is a chevron pattern. (a) $N_c=10$. 
Only \(\tau_{\mathrm{s}} \approx 0\) is observed, indicating that positive and negative
time-delay domains are nearly balanced around the center of the array.
(b) $N_c = 30$. In addition to the peak at $\tau_s \approx 0$, two peaks at positive and negative values of $\tau_s$ are observed, 
with \(|\tau_s|\) slightly smaller than in the periodic-boundary case.
(c) $N_c=50$. 
The distribution contains several nonzero peaks, while additional samples are distributed around \(\tau_s=0\).
(d) $N_c=100$. 
The values of \(\tau_{\mathrm{s}}\) are distributed over a range of values, and the central peak around \(\tau_{\mathrm{s}}\approx 0\) is more pronounced than in the
case \(N_c=50\).
}
\label{fig:o-hist}
\end{figure*}

Next, we examine this tendency statistically. Figure~\ref{fig:o-hist} shows histograms of the steady-state system-averaged time delay $\tau_s$ under open boundary conditions. For all numbers of flagella ($N_c=10$, $30$, $50$, $70$, and $100$), only the chevron pattern is observed in all samples.

Figure~\ref{fig:o-hist}(a) corresponds to the case $N_c=10$, where only $\tau_s \approx 0$ is observed. It should be emphasized that this does not represent the in-phase synchronized state; rather, it indicates that the sign change of $\tau_j$ is located near the center of the flagellar array in all samples.

Figure~\ref{fig:o-hist}(b) shows the case of $N_c = 30$. In addition to the peak at $\tau_s \approx 0$, two peaks appear at $\tau_s \approx \pm 0.038$.
The histogram resembles the periodic-boundary case  [Fig.~\ref{fig:p-hist}(b)] in that nonzero values of 
\(\tau_s\) appear, but with slightly smaller values of $|\tau_s|$.
This reflects the imbalance between positive and negative phase-difference domains in a chevron pattern, rather than a spatially uniform phase-lag state as observed under periodic boundary conditions.
Figure~\ref{fig:o-hist}(c) shows the case of $N_c = 50$. 
The distribution contains several nonzero peaks, while additional samples are distributed around $\tau_s=0$.
Figure~\ref{fig:o-hist}(d) corresponds to the case $N_c=100$. 
The values of \(\tau_s\) are distributed over a range of values, and the central peak
around \(\tau_s \approx 0\) is more pronounced than in the case \(N_c=50\).

\section{PHASE-REDUCED DESCRIPTION AND
LINEAR STABILITY ANALYSIS}
\subsection{Phase reduction}
\label{subsec:phase_reduction}

In this section, we derive a phase-reduced description for a one-dimensional array of hydrodynamically coupled flagella and investigate the stability of synchronized states. The phase-reduction procedure itself follows the Floquet-based framework developed by Kawamura and Tsubaki~\cite{kawamura2018phase} for two coupled flagella. Here, we extend their formulation to a spatially distributed one-dimensional array and derive the corresponding nearest-neighbor phase interaction model.

When the hydrodynamic interaction term in Eq.~\eqref{eq:multi-flagella-a} is sufficiently small, it can be treated as a weak perturbation. In this limit, the waveform of each flagellum is approximately identical to that of an isolated flagellum. We therefore consider the dynamics of an isolated flagellum,
\begin{equation}
\frac{\partial h}{\partial t} = \mathcal{N}(h),
\label{eq:single-flagella-main}
\end{equation}
\begin{equation}
\mathcal{N}(h) = -c \frac{\partial h}{\partial \xi} - 2 \frac{\partial^2 h}{\partial \xi^2} - \frac{\partial^4 h}{\partial \xi^4} + \left( \frac{\partial^2 h}{\partial \xi^2} \right)^3
\end{equation}
which admits a stable periodic solution. In the periodic steady state, the solution can be parameterized by a phase variable $\theta(t)$ as
\begin{equation}
h(\xi,t) = h_0(\xi,\theta(t)),
\end{equation}
\begin{equation}
\dot{\theta} = \omega,
\end{equation}
where $\omega$ is the intrinsic angular frequency and $h_0(\xi,\theta)$ satisfies
\begin{equation}
\omega \partial_\theta h_0 = \mathcal{N}(h_0), \qquad h_0(\xi,\theta+2\pi)=h_0(\xi,\theta).
\label{eq:steady-state-main}
\end{equation}

Linearization around the periodic solution yields a Floquet-type operator $L(\xi,\theta)$ (see Appendix A for details). Because of temporal translation symmetry, the operator possesses a Floquet zero eigenfunction associated with the phase mode. Differentiating Eq.~(\ref{eq:steady-state-main})
with respect to $\theta$ gives
\begin{equation}
L(\xi,\theta)\,\partial_\theta h_0 = 0,
\end{equation}
indicating that
\begin{equation}
U_0(\xi,\theta) = \partial_\theta h_0(\xi,\theta)
\end{equation}
is the zero eigenfunction.

We denote the corresponding adjoint zero eigenfunction by $U_0^*(\xi,\theta)$, satisfying
\begin{equation}
L^*(\xi,\theta)U_0^*(\xi,\theta) = 0,
\end{equation}
with normalization
\begin{equation}
\int_0^L d\xi \, U_0^*(\xi,\theta) U_0(\xi,\theta) = 1.
\label{eq:normalization}
\end{equation}
Here, $L^*$ is the adjoint of $L$ with respect to the inner product in $\xi$ and $\theta$ (see Appendix A). Following Ref.~\cite{kawamura2018phase}, this normalization is equivalent to the $2\pi$-averaged inner product
\begin{equation}
\frac{1}{2\pi}\int_0^{2\pi} d\theta \int_0^L d\xi \,
U_0(\xi,\theta)U_0^*(\xi,\theta) = 1,
\label{eq:normalization_2}
\end{equation}
and can be shown to reduce to Eq.~\eqref{eq:normalization}.

The phase dynamics of weakly coupled flagella can then be obtained by projecting the perturbation onto the phase direction using $U_0^*(\xi,\theta)$. Substituting $h_j(\xi,t)=h_0(\xi,\theta_j(t))$ into Eq.~\eqref{eq:multi-flagella-a}, we obtain
\begin{flalign}
\dot{\theta}_j \partial_\theta h_0(\xi,\theta_j)
&=
\mathcal{N}(h_0(\xi,\theta_j)) \notag\\
&\quad
+ \varepsilon\omega \Big(
 \partial_\theta h_0(\xi,\theta_{j-1})
+
 \partial_\theta h_0(\xi,\theta_{j+1})
\Big).
\label{eq:multi-flagella-b}
\end{flalign}
Here, we have set $\dot{\theta}_j=\omega$ at the unperturbed level. 
Multiplying Eq.~\eqref{eq:multi-flagella-b}
by the adjoint zero eigenfunction $U_0^*(\xi,\theta_j)$ and integrating over $\xi$, we obtain
\begin{equation}
\dot{\theta}_j
=
\omega
+
\varepsilon\omega
\int_0^L
U_0^*(\xi,\theta_j)
\left[
U_0(\xi,\theta_{j-1})
+
U_0(\xi,\theta_{j+1})
\right]
d\xi,
\end{equation}
where we have used $\mathcal{N}(h_0)=\omega U_0$ and the normalization condition~\eqref{eq:normalization}. We now define the hydrodynamic interaction function
\begin{equation}
H(\theta_j,\theta_k)
=
\int_0^L
U_0^*(\xi,\theta_j)U_0(\xi,\theta_k)\,d\xi.
\end{equation}
The phase dynamics can then be written as follows:
\begin{equation}
\dot{\theta}_j
=
\omega
+
\varepsilon\omega
\left[
H(\theta_j,\theta_{j-1})
+
H(\theta_j,\theta_{j+1})
\right].
\label{eq:phase-main}
\end{equation}

When the hydrodynamic interaction is weak ($\varepsilon \ll 1$), the phase difference $\theta_j-\theta_k$ changes only by $O(\varepsilon)$ during one oscillation period. We therefore average the interaction term over one period. We thus introduce the phase-coupling function $\Gamma(\theta_j-\theta_k)$ defined by the time average
\begin{equation}
\Gamma(\theta_j-\theta_k)
=
\frac{1}{2\pi}
\int_0^{2\pi}
d\lambda \,
H(\lambda+\theta_j,\lambda+\theta_k).
\end{equation}
Rewriting Eq.~(\ref{eq:phase-main}) using the phase-coupling function $\Gamma$, we obtain
\begin{equation}
\dot{\theta}_j
=
\omega
+
\varepsilon\omega
\left[
\Gamma(\theta_j-\theta_{j-1})
+
\Gamma(\theta_j-\theta_{j+1})
\right].
\label{eq:phase-main-2}
\end{equation}
Rewriting Eq.~(\ref{eq:phase-main-2}) in terms of the phase difference $\psi_j=\theta_{j+1}-\theta_j$, we obtain
\begin{equation}
\dot{\psi}_j(t)
=
\varepsilon\omega
\left\{
\Gamma(\psi_j)
+
\Gamma(-\psi_{j+1})
-
\Gamma(-\psi_j)
-
\Gamma(\psi_{j-1})
\right\}.
\label{eq:psi-evolution}
\end{equation}
Equation~\eqref{eq:psi-evolution} describes the time evolution of the phase difference $\psi_j$ in a one-dimensional array of hydrodynamically coupled flagella. This equation extends the two-flagella phase-reduction formulation of Ref.~\cite{kawamura2018phase} to a spatially distributed system with nearest-neighbor interactions.


\subsection{Linear stability analysis}
\label{subsec:linear_stability}

In this subsection, we examine the linear stability of the synchronized states observed in the numerical simulations under periodic boundary conditions, namely, the in-phase synchronized state and the metachronal-wave states.

Kawamura and Tsubaki~\cite{kawamura2018phase} numerically evaluated the coupling function $\Gamma(\psi)$ and showed that, for two eukaryotic flagella, only the in-phase synchronized state is stable. They also reported that the coupling function is remarkably close to a sinusoidal function. For analytical tractability, we approximate it by
\begin{equation}
\Gamma(\psi)= -A\sin(\psi+\alpha),
\label{eq:sin-approx}
\end{equation}
where $A>0$ is the coupling amplitude and $\alpha$ represents the phase shift of the coupling function.
The plotted coupling function in Ref.~\cite{kawamura2018phase} suggests a finite negative phase shift, roughly corresponding to $\alpha \simeq -0.3\pi$.
We therefore assume $-\pi/2<\alpha<0$.
Under this approximation, we show that, in a one-dimensional flagellar array, not only the in-phase synchronized state but also metachronal-wave states can become linearly stable. This result illustrates how spatial extension can qualitatively change the stability structure of synchronized states.
Substituting $\Gamma(\psi)=-A\sin(\psi+\alpha)$ into Eq.~\eqref{eq:psi-evolution}, we obtain
\begin{align}
\dot{\psi}_j(t)
= -2A\varepsilon \omega \bigg[
\cos\alpha\left(
\sin\psi_j
- \sin\psi_+
\cos\psi_-
\right)
\notag \\
-\sin\alpha
\sin\psi_+
\sin\psi_-
\bigg].
\label{eq:theta-difference-evolution-b}
\end{align}
Here we define
\begin{equation}
\psi_+ =
\frac{\psi_{j+1}+\psi_{j-1}}{2},
\qquad
\psi_- =
\frac{\psi_{j+1}-\psi_{j-1}}{2}.
\end{equation}
For a spatially uniform phase-difference state,
\begin{equation}
\psi_j = \psi_0
\qquad
\text{for all } j,
\label{eq:psi_0}
\end{equation}
substitution into Eq.~\eqref{eq:theta-difference-evolution-b} shows that the right-hand side vanishes. Hence,
\begin{equation}
\dot{\psi}_j = 0,
\end{equation}
showing that any constant phase difference constitutes a stationary solution, including the in-phase state ($\psi_0=0$) and metachronal-wave states ($\psi_0 \neq 0$).

Next, we investigate the stability of these synchronized states by performing a linear stability analysis. We introduce a perturbation around the state with a constant phase difference and substitute
\begin{equation}
\psi_j
=
\psi_0
+
\sum_k
B_k
\exp
\left(
i\frac{2\pi k}{N_c}j + \Lambda_k t
\right)
\end{equation}
into the equation. Retaining terms up to first order in $B_k$, we obtain the linear growth rate
\begin{align}
\Lambda_k &=
-2A\varepsilon\omega
\Big[
\cos\psi_0\cos\alpha
\left(1-\cos\frac{2\pi k}{N_c}\right)
\notag \\
&\quad
-i\sin\psi_0\sin\alpha
\sin\frac{2\pi k}{N_c}
\Big].
\label{eq:linear-growth-rate-a}
\end{align}
For $\cos\alpha>0$, the real part of $\Lambda_k$ becomes negative when $\cos\psi_0>0$,
indicating that the synchronized state is linearly stable. Therefore, the stability condition for the synchronized state is
\begin{equation}
-\frac{\pi}{2}<\psi_0<\frac{\pi}{2}.
\label{eq:metachronal-stability}
\end{equation}

Under periodic boundary conditions, the phase difference satisfies
\begin{equation}
\psi_0 =
\frac{2\pi n}{N_c},
\end{equation}
and the stability condition becomes
\begin{equation}
-\frac{N_c}{4} < n < \frac{N_c}{4}.
\label{eq:metachronal-stability-b} 
\end{equation}
Here, $n$ denotes an arbitrary integer. The case $n=0$ corresponds to the in-phase synchronized state, whereas $n \neq 0$ represents metachronal-wave states. As the number of flagella $N_c$ increases, the range of integers $n$ satisfying Eq.~\eqref{eq:metachronal-stability-b} expands. Thus, the number of linearly stable metachronal-wave states increases with system size, whereas the in-phase state remains a single stable mode. This multiplicity explains why metachronal-wave states are frequently selected in the numerical simulations. Since $-\pi/2<\alpha<0$ implies $\cos\alpha>0$, this stability condition holds throughout the assumed range of $\alpha$.

Because the simulations are performed at \(\varepsilon=0.4\), the weak-coupling
phase reduction should not be regarded as quantitatively exact. Rather, the
phase-reduced model is used here as a minimal analytical description that
captures the stability mechanism and the increase in the number of stable
phase-locked modes with system size.

\subsection{Continuum limit and advection--diffusion dynamics}
\label{subsec:continuum_limit}

To gain an intuitive understanding of the dynamics of the synchronized states, we analyze the long-wavelength behavior of the phase-difference dynamics. For this purpose, we take the continuum limit of Eq.~\eqref{eq:theta-difference-evolution-b} with respect to the phase difference $\psi_j$, assuming that the phase difference varies on spatial scales much larger than the inter-flagellar spacing $(\partial_\nu \psi \ll \psi/d)$.

Using the nondimensional spacing $d$ between neighboring flagella, we introduce a continuous spatial coordinate
\begin{equation}
\nu = jd.
\end{equation}
In the continuum limit, the phase difference $\psi_j$ is regarded as a continuous field,
\begin{equation}
\psi_j(t) \to \psi(\nu,t).
\end{equation}
Assuming that the phase difference varies slowly along the array, we expand $\psi_{j\pm1}$ around $\nu$ as
\begin{equation}
\psi_{j\pm1}
=
\psi(\nu\pm d,t)
=
\psi
\pm d\,\partial_\nu \psi
+
\frac{d^2}{2}\partial_\nu^2\psi
+
O(d^3).
\end{equation}
The $O(d^3)$ terms become negligible in the long-wavelength limit. Substituting these expressions into Eq.~\eqref{eq:theta-difference-evolution-b} and retaining terms up to $O(d^2)$, we obtain the following continuum equation for the phase-difference field $\psi(\nu,t)$:
\begin{equation}
\partial_t \psi
=
- V(\psi)\,\partial_\nu \psi
+ D(\psi)\,\partial_\nu^2 \psi
+ S(\psi)\,(\partial_\nu \psi)^2
.
\label{eq:psi-continuum}
\end{equation}
The first term on the right-hand side represents advection of the phase difference with a phase-dependent velocity $V(\psi)$. The second term, $D(\psi)\partial_\nu^2\psi$, corresponds to diffusion of the phase difference along the array and tends to smooth out spatial variations of $\psi$. The third term is a nonlinear gradient term that becomes relevant when the spatial variation of the phase difference is large. The coefficients $V(\psi)$, $D(\psi)$, and $S(\psi)$ are given by
\begin{align}
V(\psi) &= -2A\varepsilon\omega d\sin\alpha\sin\psi\\
D(\psi) &= A\varepsilon\omega d^2\cos\alpha\cos\psi\\
S(\psi) &= -A\varepsilon\omega d^2\cos\alpha\sin\psi
\end{align}

Since the advection velocity $V(\psi)$ depends on the phase difference, regions with different values of $\psi$ propagate at different speeds. This provides a natural explanation for the stripe patterns observed in Fig.~\ref{fig:synchronized}(d) and Fig.~\ref{fig:open}(b). Moreover, for the assumed negative phase shift $-\pi/2<\alpha<0$, $V(\psi)$ has the same sign as $\psi$ for $-\pi/2<\psi<\pi/2$. 
Thus, domains of positive phase difference are advected in the positive $\nu$ direction, whereas domains of negative phase difference are advected in the negative $\nu$ direction, as observed in Fig.~\ref{fig:open}(b).

The diffusion term provides the corresponding stability mechanism. Since $-\pi/2<\alpha<0$ implies $\cos\alpha>0$, $D(\psi)$ is positive for $-\pi/2<\psi<\pi/2$, so phase fluctuations are diffusively relaxed in this range. This recovers the stability condition for the synchronized state obtained from the linear stability analysis, Eq.~\eqref{eq:metachronal-stability}.

To make a direct comparison with the linear stability analysis, we introduce a small perturbation $\eta(\nu,t)$ around the uniform phase-difference state $\psi_0$:
\begin{equation}
\psi(\nu,t)=\psi_0+\eta(\nu,t).
\end{equation}
Expanding Eq.~(\ref{eq:psi-continuum}) to first order in $\eta$, we obtain
the advection--diffusion equation,
\begin{equation}
\partial_t \eta
=
- V_0 \partial_\nu \eta
+ D_0 \partial_\nu^2 \eta,
\end{equation}
where \(V_0=V(\psi_0)\) and \(D_0=D(\psi_0)\).
The advective term represents the propagation of phase perturbations,
whereas the diffusive term describes their relaxation along the array. Thus, in the long-wavelength limit, the collective dynamics can be interpreted as the transport and diffusion of phase disturbances around the steady metachronal state. The synchronized state is stable when
$D_0>0$, or equivalently when $\cos\alpha\cos\psi_0>0$, in agreement with the stability condition predicted from the linear growth rate~\eqref{eq:linear-growth-rate-a}.

\subsection{Propagation velocity of metachronal waves}

In this subsection, we derive the propagation velocity of the metachronal wave based on the phase-reduced description. We start from the phase equation~(\ref{eq:phase-main-2}) with the sinusoidal approximation of the coupling function~(\ref{eq:sin-approx}).

We consider a metachronal-wave solution with a constant phase difference,
\begin{equation}
\theta_j(t) = \Omega t + j \psi_0,
\label{eq:metachronal-ansatz}
\end{equation}
where $\psi_0$ is the phase difference between neighboring flagella and $\Omega$ is the collective frequency. Substituting Eq.~(\ref{eq:metachronal-ansatz}) into Eq.~(\ref{eq:phase-main-2}), we have
\begin{align}
\dot{\theta}_j
&=
\Omega \notag \\
&=
\omega
+
\varepsilon \omega
\left[
\Gamma(\psi_0)
+
\Gamma(-\psi_0)
\right].
\end{align}
Therefore, the collective frequency is given by
\begin{equation}
\Omega
=
\omega
-
2A \varepsilon \omega \sin\alpha \cos\psi_0.
\label{eq:collective-frequency}
\end{equation}

Since the phase difference between neighboring flagella is constant, the metachronal wave propagates along the array with a constant phase velocity. In the continuum limit, the phase field $\theta(\nu, t)$ is given by
\begin{equation}
\theta(\nu, t) = \Omega t + \frac{\psi_0 \nu}{d},
\label{eq:metachronal-ansatz-2}
\end{equation}
The propagation velocity can be obtained from the condition that the phase is constant along the moving frame, i.e.,
\begin{equation}
\frac{d}{dt}\theta(\nu(t),t)
=
\partial_t \theta + \frac{d\nu}{dt}\partial_\nu \theta = 0.
\end{equation}
Substituting Eq.~(\ref{eq:metachronal-ansatz-2}) yields the propagation velocity of the metachronal wave, $v_{\mathrm{m}} = d\nu/dt$, as
\begin{equation}
v_{\mathrm{m}} = -\frac{\Omega}{\partial_\nu \theta}
= -\frac{d\Omega}{\psi_0}.
\label{eq:phase-velocity-1}
\end{equation}
By inserting Eq.~(\ref{eq:collective-frequency}) into Eq.~(\ref{eq:phase-velocity-1}), we obtain
\begin{equation}
v_{\mathrm{m}}
=
-\frac{d}{\psi_0}
\left(
\omega
-
2A \varepsilon \omega \sin\alpha \cos\psi_0
\right).
\label{eq:phase-velocity-2}
\end{equation}
This equation indicates that the propagation velocity consists of a dominant contribution from the intrinsic oscillation and a correction arising from hydrodynamic interactions. The direction of propagation is determined by the sign of $\psi_0$. 
For $\psi_0 > 0$, the wave propagates in the negative $\nu$ direction, whereas for $\psi_0 < 0$, it propagates in the opposite direction. This is consistent with our numerical simulations.

\section{DISCUSSION}

In this study, we investigated synchronization and metachronal-wave formation in a one-dimensional array of hydrodynamically coupled eukaryotic flagella by combining direct numerical simulations and a phase-reduced theoretical description. We showed that, in contrast to the two-flagellum system, where only in-phase synchronization is stable, the extension from a pair to a one-dimensional array qualitatively changes the stability structure, allowing metachronal-wave states with finite phase differences to become stable.
Furthermore, the probability of observing metachronal waves increases with the number of flagella, indicating that system size plays a crucial role in the selection of collective states.
Unlike wave selection driven by tilt-induced geometric misalignment that breaks the left-right symmetry~\cite{jung2025emergence}, the metachronal-wave states found here arise through spontaneous symmetry breaking.

The key mechanism underlying this behavior can be understood from the phase-reduced description.
The sinusoidal approximation for the phase-coupling function with the phase shift $\alpha$ allows a simpler interpretation of the dynamics.
The linear stability analysis revealed that phase-locked states become stable when $\cos \psi_0 > 0$, which corresponds to the condition that long-wavelength perturbations are damped.
Furthermore, as the system size increases, the number of allowed phase-locked modes satisfying the stability condition increases, which explains why metachronal waves are more likely to be selected in larger systems.
For periodic boundary conditions, these results are consistent with previous studies of the nearest-neighbor Kuramoto model on a ring, where \(q\)-twisted states with discretized phase differences $2\pi q/N$ are selected with an approximately Gaussian probability distribution, $P(q)\propto e^{-kq^2}$, whose width grows with the system size~\cite{wiley2006size,groisman2025size}.
Related basin-selection problems have also been studied for phase-lagged nonlocal oscillator arrays~\cite{li2022effect}.
Because the sinusoidal approximation places the present phase-reduced model in this class, an important future direction is to examine basin selection using the full, generally nonsinusoidal hydrodynamic coupling function.

Our results highlight a qualitative difference between few-body and many-body systems. While the phase-coupling function obtained for two flagella favors only in-phase synchronization, extending the system to many interacting flagella reveals additional stable states that are absent in the two-body case. This indicates that, even when interactions are limited to nearest neighbors as in the present model, the collective dynamics of many-body systems cannot be fully understood from the behavior of an isolated pair, and must instead be analyzed at the level of the extended system.

The continuum description further provides insight into the dynamical processes leading to synchronization. The advection term in the phase equation indicates that regions with different phase differences propagate along the array with phase-dependent velocities, which is consistent with the stripe patterns observed in the transient dynamics. 
Moreover, the antisymmetry of the advection velocity $V(\psi)$ with respect to \(\psi\)
provides a natural interpretation of the chevron patterns observed under open
boundary conditions, where domains with opposite phase differences propagate in
opposite directions and meet near a stationary interface.
The stability of phase-locked states is controlled by the diffusive relaxation of phase fluctuations, characterized by $D_0>0$, which provides a clear physical interpretation of the numerical results.

Despite these insights, the present model involves several simplifying assumptions. First, we considered only nearest-neighbor hydrodynamic interactions, whereas real flagellar arrays exhibit long-range interactions mediated by the fluid.
Long-range hydrodynamic interactions can qualitatively alter collective phase-locking and pattern selection in extended oscillator arrays, leading to traveling waves, chevron patterns, phase defects, and other many-body dynamical states~\cite{brumley2016long,uchida2010synchronization}.
Second, the phase-coupling function was approximated by a sinusoidal form for analytical tractability. Although this approximation captures the qualitative features reported in previous studies, it may not accurately reproduce the detailed shape of the coupling function obtained from full numerical evaluation, and could affect the quantitative prediction of stability boundaries. Third, the analysis was restricted to a one-dimensional geometry, while many biological systems involve two-dimensional ciliary carpets. These limitations may affect quantitative aspects of the results and should be addressed in future work.

Future extensions of this study include incorporating long-range hydrodynamic interactions, exploring higher-dimensional arrays, and quantitatively comparing the theoretical predictions with experimental observations of ciliary and flagellar systems. In particular, understanding how geometric constraints and boundary conditions influence the selection and robustness of metachronal waves will be an important step toward bridging the gap between minimal theoretical models and biological systems. Such comparisons may also help connect the present theory to biomimetic platforms and artificial ciliary systems, where programmable metachronal motion and its effect on fluid transport have recently been investigated experimentally~\cite{wang2024programmable}. It will also be interesting to go beyond the steady Stokes-like description adopted here, since recent work has shown that unsteady viscous flow and vorticity diffusion can qualitatively modify synchronization and metachronal-wave selection in ciliary chains~\cite{kenne2025synchronization}.

In summary, we have shown that spatial extension and hydrodynamic coupling cooperatively give rise to stable metachronal waves in flagellar arrays. 
The present work thus connects microscopic flagellar dynamics, phase synchronization, and macroscopic wave propagation within a minimal one-dimensional setting.

\section*{Data and Code Availability}
The source code, numerical data, and supplementary movies associated with this
study are publicly available in Zenodo~\cite{zenodo_data}.

\begin{acknowledgments}
Y.W. acknowledges support from the GP-MS at Tohoku University.
This work was supported by JSPS KAKENHI Grant Number JP24K06895 to N.U.
\end{acknowledgments}

\appendix
\section{Floquet Operator and Adjoint Problem}

We follow the Floquet-based phase-reduction framework of Ref.~\cite{kawamura2018phase} and present the explicit forms of the linear and adjoint operators used in the present model.

Linearizing Eq.~(\ref{eq:single-flagella-main}) around the periodic solution $h_0(\xi,\theta)$ defined in Eq.~(\ref{eq:steady-state-main}), we introduce a perturbation
\begin{equation}
h(\xi,t)=h_0(\xi,\theta(t))+u(\xi,\theta(t),t).
\end{equation}
Substituting this into the governing equation and retaining terms linear in $u$, we obtain
\begin{equation}
\frac{\partial}{\partial t}u(\xi,\theta,t)=L(\xi,\theta)u(\xi,\theta,t),
\end{equation}
where
\begin{equation}
L(\xi,\theta)=J(\xi,\theta)-\omega\frac{\partial}{\partial\theta},
\end{equation}
\begin{equation}
J(\xi,\theta)u
=
-c\frac{\partial u}{\partial \xi}
-2\frac{\partial^2 u}{\partial \xi^2}
-\frac{\partial^4 u}{\partial \xi^4}
+
3\left(
\frac{\partial^2 h_0}{\partial \xi^2}
\right)^2
\frac{\partial^2 u}{\partial \xi^2}.
\end{equation}
Since $L(\xi,\theta)$ is $2\pi$-periodic in $\theta$, the linearized equation constitutes a Floquet-type system.

To define the adjoint operator, we introduce the inner product
\begin{equation}
[[u,u^*]]
=
\frac{1}{2\pi}
\int_0^{2\pi} d\theta \int_0^L d\xi \,
u(\xi,\theta)\,u^*(\xi,\theta),
\end{equation}
where $u^*(\xi,\theta)$ denotes the adjoint function on which the adjoint operator $L^*$ acts. The adjoint operator $L^*$ is defined by
\begin{equation}
[[u^*,Lu]]=[[L^*u^*,u]].
\end{equation}
With respect to this inner product, the adjoint operators are given by
\begin{equation}
J^*(\xi,\theta)u^*
=
c\frac{\partial u^*}{\partial \xi}
-2\frac{\partial^2 u^*}{\partial \xi^2}
-\frac{\partial^4 u^*}{\partial \xi^4}
+
3\frac{\partial^2}{\partial \xi^2}
\left[
u^*
\left(
\frac{\partial^2 h_0}{\partial \xi^2}
\right)^2
\right],
\end{equation}
\begin{equation}
L^*(\xi,\theta)=J^*(\xi,\theta)+\omega\frac{\partial}{\partial\theta}.
\end{equation}

\bibliographystyle{apsrev4-2}

%

\end{document}